\documentclass[twocolumn,preprintnumbers,a4paper,
superscriptaddress,floatfix,twoside]{revtex4}

\usepackage{bm}
\usepackage{epsfig}
\usepackage{graphics}
\usepackage{natbib}

\usepackage{fancyh}
\usepackage[l]{floatflt}
\usepackage{epsfig}
\usepackage{amssymb}
\usepackage{latexsym}
\usepackage{times}
\usepackage{slashed}
\usepackage{upgreek}
\usepackage{amsmath}
\usepackage{amsfonts}
\usepackage{amsbsy}
\usepackage{amscd}
\usepackage{bbm}

\newcommand{\eec}{\end{center}}
\newcommand{\bec}{\begin{center}}

\newcommand{\eem}{\end{matrix}}
\newcommand{\bem}{\begin{matrix}}
\newcommand{\eeq}{\end{equation}}
\newcommand{\beq}{\begin{equation}}
\newcommand{\ba}{\begin{array}}
\newcommand{\ea}{\end{array}}
\newcommand{\bea}{\begin{eqnarray}}
\newcommand{\eea}{\end{eqnarray}}
\newcommand{\baq}{\begin{eqnarray}}
\newcommand{\eaq}{\end{eqnarray}}
\newcommand{\beqs}{\begin{subequations}}
\newcommand{\eeqs}{\end{subequations}}
\def\th{{s}}

\newcommand\eqs[2]{Eqs.~(\ref{#1}) and (\ref{#2})}

\newcommand\eqss[3]{Eqs.~(\ref{#1}), (\ref{#2}) and (\ref{#3})}

\newcommand{\ftn}{\footnotesize}

\newcommand{\GeV}{{\mbox{\rm GeV}}}

\newcommand{\sFref}[2]{Fig.~\ref{#1}-{\ftn \sf ({#2})}}

\newcommand{\etal}{{\it et al.\/}}

\def\to{\rightarrow}

\def\lf{\left(}
\def\rg{\right)}
\newcommand\vev[1]{\langle {#1} \rangle}

\newcommand{\Ns}{\ensuremath{N_{\star}}}
\newcommand{\Gsm}{\ensuremath{G_{\rm SM}}}
\newcommand{\Nr}{\ensuremath{{\sf N}}}
\newcommand{\Vhi}{\ensuremath{V_{\rm I}}}
\newcommand{\Hhi}{\ensuremath{H_{\rm I}}}
\newcommand{\Whi}{\ensuremath{W}}
\newcommand{\Khi}{\ensuremath{K}}

\newcommand{\Vhio}{\ensuremath{V_{\rm I0}}}
\newcommand{\Vtr}{\ensuremath{V_{\rm tr}}}

\newcommand{\mP}{\ensuremath{m_{\rm P}}}

\newcommand{\Mgut}{\ensuremath{M_{\rm GUT}}}
\newcommand{\Ggut}{\ensuremath{G}}
\newcommand{\Gfl}{\ensuremath{G_{\rm 5_X}}}
\newcommand{\Glr}{\ensuremath{G_{\rm LR}}}

\newcommand{\aS}{\ensuremath{{\rm a}_S}}
\newcommand{\ms}{\ensuremath{{\widetilde m}}}

\newcommand{\ns}{\ensuremath{n_{\rm s}}}
\newcommand{\As}{\ensuremath{A_{\rm s}}}
\newcommand{\as}{\ensuremath{\alpha_{\rm s}}}

\newcommand{\xsg}{\ensuremath{\sigma}}
\newcommand{\xm}{\ensuremath{M}}

\def\al{{\alpha}}

\def\bt{{\beta}}
\def\bbet{{\bar\beta}}

\newcommand{\im}{\ensuremath{{\sf Im}}}

\newcommand{\sg}{\ensuremath{\sigma}}
\newcommand{\sgex}{\ensuremath{\sigma_\star}}
\newcommand{\sgc}{\ensuremath{\sigma_{\rm c}}}
\newcommand{\sgf}{\ensuremath{\sigma_{\rm f}}}

\newcommand{\Ld}{\ensuremath{\Lambda}}
\newcommand{\kp}{\ensuremath{\kappa}}
\newcommand{\ks}{\ensuremath{k_{S}}}
\newcommand{\ksp}{\ensuremath{k_{S\Phi}}}
\newcommand{\kspb}{\ensuremath{k_{S\bar\Phi}}}

\newcommand{\hepth}[1]{{\ftn \tt hep-th/#1}}
\newcommand{\hepph}[1]{{\ftn\tt hep-ph/#1}}

\newcommand{\arxiv}[1]{{\ftn\tt  arXiv:#1}}

\renewcommand{\thesubsection}{{\small\sf\Alph{subsection}}}

\newcommand{\Eref}[1]{Eq.~(\ref{#1})}
\newcommand{\Sref}[1]{Sec.~\ref{#1}}
\newcommand{\Fref}[1]{Fig.~\ref{#1}}
\newcommand{\Tref}[1]{Table~\ref{#1}}
\newcommand{\cref}[1]{Ref.~\cite{#1}}

\def\Ka{K\"{a}hler potential~}
\def\Kap{K\"{a}hler potential}
\def\trns{trans-Planckian}
\def\sub{sub-Planckian}

\newcommand{\bdhh}{{\ensuremath{\normalsize I{\kern-2.9pt H}}}}

\renewcommand{\refname}{{\bf\scshape References}}
\newcommand{\bicep}{{\scshape Bicep2}}

\renewenvironment{subequations}{%
\refstepcounter{equation}%
\setcounter{parentequation}{\value{equation}}%
  \setcounter{equation}{0}
  \ignorespaces
}{%
  \setcounter{equation}{\value{parentequation}}%
  \ignorespacesafterend
}

\begin{document}


\title{\boldmath\bf\scshape  From Hybrid to Quadratic Inflation With High-Scale Supersymmetry Breaking}

\author{\scshape Constantinos Pallis}
\affiliation{Departament de F\'isica Te\`orica and IFIC,
Universitat de Val\`encia-CSIC, E-46100 Burjassot, SPAIN
\\  {\sl e-mail address: }{\ftn\tt cpallis@ific.uv.es}}
\author{\scshape  Qaisar Shafi}
\affiliation{ Bartol Research Institute, Department of Physics and
Astronomy, University of Delaware, Newark, DE 19716, USA\\  {\sl
e-mail address: }{\ftn\tt shafi@bartol.udel.edu}}


\begin{abstract}

\noindent {\ftn \bf\scshape Abstract:} Motivated by the reported
discovery of inflationary gravity waves by the \bicep\ experiment,
we propose an inflationary scenario in supergravity, based on the
standard superpotential used in hybrid inflation. The new model
yields a tensor-to-scalar ratio $r\simeq0.14$ and scalar spectral
index $\ns\simeq0.964$, corresponding to quadratic (chaotic)
inflation. The important new ingredients are the high-scale,
$(1.6-10)\cdot10^{13}$ GeV, soft supersymmetry breaking mass for
the gauge singlet inflaton field and a shift symmetry imposed on
the \Kap. The end of inflation is accompanied, as in the earlier
hybrid inflation models, by the breaking of a gauge symmetry at
$(1.2-7.1)\cdot 10^{16}~{\rm GeV}$, comparable to the
grand-unification scale.
\\ \\ {\scriptsize {\sf PACs numbers: 98.80.Cq, 12.60.Jv}


}

\end{abstract}\pagestyle{fancyplain}

\maketitle

\rhead[\fancyplain{}{ \bf \thepage}]{\fancyplain{}{\sl From Hybrid
to Quadratic Inflation With High-Scale SUSY Breaking}}
\lhead[\fancyplain{}{\sl \leftmark}]{\fancyplain{}{\bf \thepage}}
\cfoot{}

\section{Introduction}

The discovery of B-modes in the polarization of the cosmic
microwave background radiation at large angular scales by the
\bicep\ experiment \cite{gws} has created much excitement among
inflationary model builders, since this effect can be caused by an
early inflationary era with a large tensor-to-scalar ratio $r=
0.16^{+0.06}_{-0.05}$ -- after substraction of a dust foreground.
Although other interpretations \cite{gws1,gws2} of this result are
possible, it motivates us to explore how realistic
\emph{supersymmetric} ({\sf\ftn SUSY}) inflation models can
accommodate such large $r$ values.

The textbook quadratic inflationary model \cite{chaotic}
predicting $r=0.13 - 0.16$, and a (scalar) spectral index $\ns=
0.96-0.967$, seems to be in good agreement with \bicep\ ($r$) and
the WMAP \cite{wmap} and Planck \cite{plin} measurements ($\ns$).
Quadratic inflation can be accompanied by a \emph{Grand Unified
Theory} ({\sf \ftn GUT}) phase transition in non-supersymmetric
inflation models, based either on the Coleman-Weinberg or Higgs
\cite{mqw} potential, which yield predictions for $\ns$ that more
or less overlap with the prediction of the quadratic model
\cite{oss,seto}. However, significant differences appear between
the predictions of $r$ in these models which can be settled
through precision measurements. The consistent supersymmetrization
of these models is a highly non-trivial task due to the \trns\
values of the inflaton field which aggravate the well-known
$\eta$-problem within \emph{supergravity} ({\sf\ftn SUGRA}).

One of the more elegant SUSY models which nicely combines
inflation with a GUT phase transition is the model of \emph{F-term
hybrid inflation} \cite{susyhybrid, hybrid} -- referred as {\sf
\ftn FHI}. It is based on a unique renormalizable superpotential,
dictated by a $U(1)$ R-symmetry, employs \sub\ values for the
inflaton field and can be naturally followed by the breaking of a
GUT gauge symmetry, $\Ggut$, such as $G_{B-L}= G_{\rm SM}\times
U(1)_{B-L}$ \cite{mfhi} -- where ${G_{\rm SM}}= SU(3)_{\rm
C}\times SU(2)_{\rm L}\times U(1)_{Y}$ is the gauge group of the
\emph{Standard Model} ({\ftn\sf SM}) -- $\Glr=SU(3)_{\rm C}\times
SU(2)_{\rm L} \times SU(2)_{\rm R} \times U(1)_{B-L}$
\cite{dvali}, and flipped $SU(5)$ \cite{flipped}, with gauge
symmetry $\Gfl=SU(5)\times U(1)_X$. The embedding of the simplest
model of FHI within a GUT based on a higher gauge group may suffer
from the production of disastrous cosmic defects which can be
evaded, though, by using shifted \cite{shifted}  or smooth
\cite{smooth} FHI.

In the simplest realization of FHI the standard \cite{susyhybrid}
superpotential is accompanied by a minimal (or canonical) \Kap.
The resulting $\ns$ is found to be in good agreement with the WMAP
and Planck data after including in the inflationary potential
\emph{radiative corrections} ({\sf \small RCs}) \cite{susyhybrid}
and the \emph{soft SUSY breaking} ({\sf\ftn SSB}) linear term
\cite{sstad, mfhi} -- with a mass parameter in the TeV range -- a
SSB mass term for the inflaton in the same energy region can be
ignored in this analysis. This scenario yields \cite{mfhi} $r$
values which lie many orders of magnitude below the measurement
reported \cite{gws} by \bicep. A more elaborate extension of this
standard FHI scenario exploits non-minimal, quasi-canonical \Kap s
\cite{gpp,rlarge} or SSB mass of magnitude as large as
$10^{10}~\GeV$ for the inflaton field \cite{rlarge1}. Depending on
the underlying assumptions, the predictions for $r$ are
considerably enhanced compared to the minimal scenario of
\cref{sstad, mfhi}. Thus, $r$ values as large as $0.01$ to $0.03$
have been reported \cite{rlarge1, rlarge}; this fact certainly
puts $r$ in the observable range, but it still remains an order of
magnitude below the \bicep\ measurement -- however, see
\cref{domke} for models of FHI with \Ka not-respecting the
R-symmetry.

Motivated by this apparent discrepancy between the large $r$
values reported by \bicep\ and the predictions of FHI models, we
present here a modified scenario of F-term inflation in which a
quadratic potential dictates the inflationary phase, thus
replicating the predictions of quadratic inflation, employing the
well-studied standard superpotential of FHI. The two key elements
for successfully implementing this scenario include a judicious
choice of the \Ka and a high-scale SUSY breaking. In particular,
following earlier similar attempts \cite{shift1} a shift symmetry
is imposed on the \Ka to protect the inflationary potential from
SUGRA corrections which are dangerous due to \trns\ inflaton field
values. Moreover, we assume that SUSY is broken at an intermediate
scale, $\ms\sim 10^{13}~\GeV$, which can be identified with the
SSB mass of the inflaton. In the context of high-scale SUSY
\cite{strumia,ibanez}, such a large SSB scale can become
consistent with the LHC results \cite{lhc} on the mass,
$m_h\simeq126~\GeV$, of the SM Higgs boson, $h$. The end of
inflation can be accompanied by the breaking of some gauge
symmetry such as $\Glr$ or $\Gfl$ with the gauge symmetry breaking
scale $M$ assuming values close to the SUSY GUT scale
$\Mgut\simeq2.86\cdot10^{16}~{\rm GeV}$.

Below, we describe in \Sref{fhi} the basic ingredients of our
inflationary scenario. Employing a number of constraints presented
in \Sref{fhi4}, we provide restrictions on the model parameters in
\Sref{res}. Our conclusions are summarized in Sec.~\ref{con}.
Henceforth we use units where the reduced Planck scale $\mP =
2.44\cdot 10^{18}~\GeV$ is taken equal to unity.

\section{The Inflationary Scenario}\label{fhi}

\subsection{\sf\scshape\small The GUT Symmetry
Breaking}\label{fhi1} In the standard FHI we adopt the
superpotential
\beq W = \kappa S\left(\bar \Phi\Phi-M^2\right),\label{Whi}\eeq
which is the most general renormalizable superpotential consistent
with a continuous R-symmetry \cite{susyhybrid} under which \beq S\
\to\ e^{i\alpha}\,S,~\bar\Phi\Phi\ \to\ \bar\Phi\Phi,~W \to\
e^{i\alpha}\, W.\label{Rsym} \eeq Here $S$ is a $\Ggut$-singlet
left-handed superfield, and the parameters $\kappa$ and $M$ are
made positive by field redefinitions. In our approach
$\bar{\Phi}$, $\Phi$ are identified with a pair of left-handed
superfields conjugate under $\Ggut$ which break $\Ggut$ down to
$\Gsm$. Indeed,  along the D-flat direction $|\bar\Phi|=|\Phi|$
the SUSY potential, $V_{\rm SUSY}$, extracted -- see e.g.
\cref{review} -- from $W$ in Eq.~(\ref{Whi}), reads
\beq \label{VF} V_{\rm SUSY}=
\kappa^2\left((|\Phi|^2-M^2)^2+2|S|^2 |\Phi|^2\right).\eeq
From $V_{\rm SUSY}$ in Eq.~(\ref{VF}) we find that the SUSY vacuum
lies at
\beq
|\vev{S}|=0\>\>\>\mbox{and}\>\>\>\left|\vev{\Phi}\right|=\left|\vev{\bar\Phi}\right|=M,
\label{vevs} \eeq
where the vacuum expectation values of $\Phi$ and $\bar\Phi$ lie
along their SM singlet components. As a consequence, $\Whi$ leads
to the spontaneous breaking of $\Ggut$ to $\Gsm$.

\subsection{\sf\scshape\small The Inflationary Set-up}\label{fhi2}

It is well-known \cite{susyhybrid} that $\Whi$ also gives rise to
FHI since, for values of $|S| \gg M $, there exist a flat
direction
\begin{equation} \label{inftr}\th\equiv\sqrt{2}\,\im[S]=0~~~\mbox{and}~~~\bar\Phi={\Phi}=0, \eeq
which provides us with a constant potential energy $\kp^2M^4$
suitable for supporting FHI. The inclusion of SUGRA corrections
with canonical (minimal) \Ka does not affect this result at the
lowest order in the expansion of $S$ -- due to a miraculus
cancelation occuring. The SUGRA corrections with quasi-canonical
\Ka \cite{gpp, rlarge} can be kept under control by mildly tuning
the relevant coefficients thanks to \sub\ $S$ values required by
FHI. The resulting $\ns$ values can be fully compatible with the
data \cite{wmap,plin} but the predicted $r$ \cite{rlarge1,rlarge}
remains well below the purported measurement reported by \bicep.

In order to safely implement quadratic inflation, favored by
\bicep, within SUGRA and employing $W$ in \Eref{Whi}, we have to
tame the $\eta$ problem which is more challenging due to the
\trns\ values needed for the inflaton superfield, $S$. To this
end, we exploit a \Ka\ which respects the following symmetries:
\beq S \to\ S+c \>\>\>\mbox{and}\>\>\> S \to\ -
S,\label{shift}\eeq
where $c$ is a real number -- cf.~\cref{shift1}. Namely we take
\bea\nonumber K&=&-\frac12(S-S^*)^2+|\Phi|^2+|\bar \Phi|^2
\\\nonumber  &+&{(S-S^*)^2\over2\Ld^2}\lf\ks (S-S^*)^2+\ksp |\Phi|^2+\kspb |\bar \Phi|^2\rg \\ &+&{1\over\Ld^2}\lf k_{\Phi} |\Phi|^4+k_{\bar \Phi} |\bar
\Phi|^4\rg+\cdots~~~~\cdot\label{Khi} \eea
Here $\ks, k_{\Phi}, k_{\bar \Phi}, \ksp$ and $\kspb$ are positive
or negative constants of order unity -- for simplicity we take
$\ksp=\kspb$ -- and $\Ld$ is a cutoff scale determined below.
Although $K$ is not invariant under the $R$ symmetry of
\Eref{Rsym}, the fields $\Phi^\al=S,\Phi,\bar\Phi$ are canonically
normalized, i.e., $K_{\al\bbet}=\delta_{\al\bbet}$ -- note that
the complex scalar components of the various superfields are
denoted by the same symbol.

The F--term (tree level) SUGRA scalar potential, $\Vhio$, of our
model is obtained from $W$ in Eq.~(\ref{Whi}) and $\Khi$ in
\Eref{Khi} by applying the standard formula:
\beq \Vhio=e^{\Khi}\left(K^{\al\bbet}{\rm F}_\al {\rm
F}_\bbet-3{\vert W\vert^2}\right),\label{Vsugra} \eeq with
$K_{\al\bbet}={\Khi_{,\Phi^\al\Phi^{*\bbet}}}$,
$K^{\bbet\al}K_{\al\bar \gamma}=\delta^\bbet_{\bar \gamma}$ and
${\rm F}_\al=W_{,\Phi^\al} +K_{,\Phi^\al}W.$ We explicitly verify
that the SUSY vacuum of \Eref{vevs} remains intact for the choice
of $K$ in \Eref{Khi}. Along the field direction in \Eref{inftr}
the only surviving terms of $\Vhio$ are
\beq\Vhio=e^{K}\lf K^{SS^*}\, |W_{,S}|^2-
3|W|^2\rg=\kp^2\xm^4\lf1-{3\over2}\xsg^2\rg,\label{Vhi0}\eeq
where the canonically normalized inflaton, $\sg$, is defined by
\beq  S=\lf\sg+i\th\rg /\sqrt{2}. \label{xs1}\eeq
As shown from \Eref{Vhi0}, $\Vhio$ is not suitable to drive
inflation mainly due to the minus sign which renders $\Vhio$
unbounded from below for large $\sigma$'s -- cf. \cref{senoguz}.
On the other hand, the symmetries in \Eref{shift} ensure a
complete disappearance of the exponential prefactor in
\Eref{Vhi0}, which could ruin any inflationary solution for large
$\xsg$'s.

A satisfactory solution can be achieved, if we consider an
intermediate-scale SSB mass parameter $\ms$, whose contribution
can exceed the negative contribution to $\Vhio$ for conveniently
selected $\kp$ and $\xm$. Such a heavy mass parameter is normally
generated following the usual SUSY breaking procedures -- see e.g.
\cref{nilles} -- provided that the gravitino mass is of similar
size and the Polonyi field has canonical \Kap. The contributions
to the inflationary potential from the SSB effects
\cite{sstad,mfhi} can be parameterized as follows:
\beqs\begin{equation} \label{Vis}  V_{\rm IS}=\ms^2
\mbox{$\sum_\al$}|\Phi^\al|^2 -\lf \aS\kp M^2 S-\kp A_\kp S
\Phi\bar\Phi+{\rm c.c.}\rg,
\end{equation}
where we assume for simplicity that there is a universal SSB mass
$\ms$ for all the superfields $\Phi^\al=S,\Phi,\bar\Phi$ of our
model. Also ${\rm a}_S$ and $A_\kp$ are mass parameters comparable
to $\ms$. Along the field configuration in \Eref{inftr}, $V_{\rm
IS}$ reads
\begin{equation} \label{Vis0}  V_{\rm IS}=\ms^2 \xsg^2/2-\sqrt{2} \aS \kp
\xm^2\xsg. \end{equation}\eeqs
We note in passing that, due to \Eref{Vis}, $|\vev{S}|$ is shifted
\cite{dvali} from its value in \Eref{vevs} to
\beq|\vev{S}|\simeq\lf
|A_\kp|-|\aS|\rg/2\kp(1+\ms^2/2\kp^2M^2),\label{vevn}\eeq where we
selected conveniently the phases of $A_\kp$ and $\aS$ so that
$\vev{V_{\rm SUSY}+V_{\rm IS}}$ is minimized.

\renewcommand{\arraystretch}{1.2}

\begin{table}[!t]
\caption{\normalfont The mass spectrum of the model along the path
in \Eref{inftr}.}
\begin{tabular}{c|@{\hspace{0.1cm}}c@{\hspace{0.1cm}}|@{\hspace{0.1cm}} c}\toprule
{Fields} &{Eingestates} & {Masses Squared}\\ \colrule
\multicolumn{3}{c}{Bosons}\\ \colrule
$1$ real scalar &$\sg$ & $m_{\sg}^2=\ms^2 -3\kp^2 \xm^4  $\\
$1$ real scalar &$\th$ & $m_{\th}^2=\ms^2+\kp^2 M^4 $\\
&&$\cdot\lf(3-\sg^2)-24
\ks/\Ld^{2}\rg $\\
$2{\sf N}$ complex &
$\phi_{i\pm}={\bar{\phi}_i\pm{\phi}_i\over\sqrt{2}}$ &
$m_{{\phi}_\pm}^2\simeq{\ksp \kp^2M^4 \over\Ld^{2}}\mp{\kp
|A_\kp|\sg\over\sqrt{2}}+$\\
scalars& $(i=1,2)$ & $\ms^2+\kp^2\lf{(1\pm M^2)\sg^2\over2}\mp
M^2\rg $\\\colrule
\multicolumn{3}{c}{Fermions}\\ \colrule
$1$ Weyl spinor & ${{\psi}_{S}}$& $m_{\psi_S}^2=\kp^2 \xm^2 \xsg^2/2$\\
$2{\sf N}$ Weyl spinors & ${\psi}_\pm={{\psi}_{\bar \Phi}\pm {\psi}_{\Phi}\over\sqrt{2}}$& $m_{{\psi}_\pm}^2=\kp^2 \xsg^2/2$\\
\botrule
\end{tabular}
\label{tab}
\end{table}

\subsection{\sf\scshape\small Beyond the Tree-Level
Potential}\label{fhi3} Expanding the various fields, besides $S$
-- see \Eref{xs1} --, in real and imaginary parts according to the
prescription
\beq X= {\lf x_1+ix_2\rg/\sqrt{2}}\label{cannor} \eeq
where $X=\Phi, \bar\Phi$ and $x=\phi, \bar \phi$ respectively, we
are able to check the stability of the field directions in
\Eref{inftr}. Namely, we check the validity of the conditions
\beqs \beq {\partial \Vtr/\partial\chi^\al}=0\>\>\>
\mbox{and}\>\>\> m^2_{\chi^\al}>0,  \label{Vcon} \eeq
where $\chi^\al=\sg, \th, \phi_i$ and $\bar\phi_i$ with $i=1,2$
and $\Vtr$ stands for the tree-level inflationary potential
\beq \Vtr=\Vhio+V_{\rm IS} \label{Vtr} \eeq\eeqs
with $\Vhio$ and $V_{\rm IS}$ given in Eq.~(\ref{Vhi0}) and
(\ref{Vis0}). Note that the imposed $\mathbb{Z}_2$ symmetry on $K$
-- see \Eref{shift} -- excludes the terms $(S-S^*)$ or $(S-S^*)^3$
which could violate the first condition in \Eref{Vcon} for
$\chi^\al=\th$. Moreover, in \Eref{Vcon}, $m^2_{\chi^\al}$ are the
eigenvalues of the mass squared matrix
$M^2_{\al\bt}=\partial^2\Vtr/\partial\chi^\al
\partial\chi^\beta$ which are presented in \Tref{tab}. Setting
\beqs\beq \ms\geq\sqrt{3}\kp M^2,~~\Ld\leq {2
\sqrt{3|\ks|}\over\sqrt{2\Ns-3}}\label{con1}\eeq (where we employ
\Eref{sr} and set $\aS\ll1$ for the derivation of the latter
expression above) and, neglecting $M^4$ terms, \beq\sg\geq \sgc
\simeq\frac{\sqrt{2} \sqrt{\kp^2 M^2-\ms^2}}{\kp\sqrt{1 +
M^2}}~~\mbox{with}~~M>{\ms\over \kp} \label{sgc}\eeq\eeqs
assists us to achieve the positivity of $m_{\sg}^2$, $m_{\th}^2$
and $m_{\phi+}^2$, respectively. Note that the two first terms in
the expression for $m^2_{\phi\pm}$ are neglected in the derivation
of \Eref{sgc}, since their contribution is suppressed for
$k_{S\Phi}\sim1$ and $|A_\kp|\simeq10^{-6}-10^{-5}$. In \Tref{tab}
we also present the masses squared of the chiral fermions of the
model along the trajectory in \Eref{inftr}. We remark that the
fermionic and bosonic degrees of freedom are equal to $2(1+2{\sf
N})$. Inserting these masses into the well-known Coleman-Weinberg
formula, we can find the one-loop RCs, $\Delta V$, which can be
written as
\bea \nonumber \Delta V&=& {1\over64\pi^2}\lf
m_{\sg}^4\ln{m_{\sg}^2\over Q^2}+m_{\th}^4\ln{m_{\th}^2\over Q^2}
-2m_{\psi_S}^4\ln{m_{\psi_S}^2\over Q^2}\right.\\
&+&\left.2{\sf
N}\lf\mbox{$\sum_{i=\pm}$}m_{{\phi}_i}^4\ln{m_{{\phi}_i}^2\over
Q^2} -2m_{{\psi}_\pm}^4\ln{m_{{\psi}_\pm}^2\over
Q^2}\rg\right).~~~~~~~\label{Vrc}\eea
Here $Q$ is a renormalization group mass scale and  $\Nr$ is the
dimensionality of the representations to which $\bar{\Phi}$ and
$\Phi$ belong -- we have \cite{mfhi,rlarge} $\Nr=1, 2, 10$ for
$\Ggut=G_{B-L}, \Glr$ and $\Gfl$, correspondingly.

All in all, the full potential of our model is
\beq \Vhi=\Vtr+\Delta V,\label{Vhic} \eeq
with $\Vtr$ and $\Delta V$ given in Eq.~(\ref{Vtr}) and
(\ref{Vrc}) respectively.

\section{Constraining the Model Parameters}\label{fhi4}

Based on $\Vhi$ in \Eref{Vhic} we proceed to explore the allowed
parameter space of our model employing the standard slow-roll
approximation \cite{review}. The free parameters are
$$ \kp,~M,~\ks,~\ksp,~\Ld,~\ms,~\aS,~|A_\kp|~~\mbox{and}~~\Nr\,.$$
The parameters $\ks, \ksp$ and $|A_\kp|$ exclusively influence the
values of $m^2_{\th}$ and $m^2_{\phi\pm}$ -- see \Tref{tab} -- and
so, we take for them a convenient value, close to unity, which can
assist us to achieve the positivity and heaviness -- see below --
of these masses squared, e.g., $\ks=-\ksp=-5$ and
$|A_\kp|=10^{-6}$. The remaining parameters can be restricted by
imposing a number of observational (1,3) and theoretical (2)
restrictions specified below:

\subsection{\sf\scshape\small Inflationary
Observables}\label{fhi4a} The number of e-foldings, $\Ns$, that
the pivot scale $k_\star=0.05/{\rm Mpc}$ undergoes during
inflation, and the amplitude $A_{\rm s}$ of the power spectrum of
the curvature perturbation can be calculated using the standard
formulae
\begin{equation}  \label{Nhi}
N_{\star}=\int_{\sigma_{\rm f}}^{\sigma_{\star}}\, {d\sigma}\:
\frac{V_{\rm I}}{V'_{\rm I}}~~\mbox{and}~~\sqrt{A_{\rm s}}=
\frac{1}{2\sqrt{3}\, \pi}\; \left.\frac{V_{\rm
I}^{3/2}(\sigma_\star)}{|V'_{{\rm I}}(\sigma_\star) |}\right.
\end{equation}
where the prime denotes derivation with respect to $\sigma$,
$\sigma_{\star}$ is the value of $\sigma$ when $k_\star$ crosses
outside the horizon of inflation, and $\sigma_{\rm f}$ is the
value of $\sigma$ at the end of inflation which coincides with
$\sgc$, \Eref{sgc}, if $\epsilon(\sgc)\leq1$ and $\eta(\sgc)\leq1$
or is determined by the condition:
\beqs\beq {\ftn\sf
max}\{\epsilon(\sigma),\eta(\sigma)\}=1~~\mbox{for}~~\sg\geq\sgc.\label{slc}\eeq
Here $\epsilon$ and $\eta$ are the well-known \cite{review}
slow-roll parameters defined as follows:
\beq \label{slow} \epsilon=\left({V'_{\rm I}}/\sqrt{2}{V_{\rm
I}}\right)^2~~\mbox{and}~~\eta={V''_{\rm I}}/{V_{\rm I}}.
\eeq\eeqs
Agreement with the observations \cite{wmap, plin} requires
\begin{equation} \label{Prob}
N_{\star}\simeq55~~\mbox{and}~~\sqrt{A_{\rm s}}\simeq\: 4.686\cdot
10^{-5},
\end{equation}
which allow us to restrict  $\sigma_\star$ and $\ms$. Neglecting
$\Delta V$ in \Eref{Vhic} and assuming that $\aS$ is adequately
suppressed we approach the quadratic inflationary model with
\beqs\beq
\epsilon=\eta=2/\sg^2,~~\sgf\simeq\sqrt{2}~~\mbox{and}~~\sgex\simeq2\sqrt{\Ns}.\label{sr}\eeq
Hence, inflation takes place for $\sg\gg1$ with $\sgf\sim1$ and
$\sgc\ll1$ -- see \Eref{sgc}. Employing the last equalities in
\eqs{Nhi}{sr} we find
\beq \ms\simeq\sqrt{3} \sqrt{\kp^2 M^4\Ns^2 + 2 \As
\pi^2}/\Ns=(6-40)\cdot10^{-6},\label{lan}\eeq\eeqs
for the values of \Eref{Prob} and $\kp$ and $M$ of order 0.01.
Therefore, the range of the $\ms$ values is somehow extended
compared to those obtained in the quadratic model.

We can finally calculate $\ns$, its running, $\as$, and $r$, via
the relations:
\beqs\bea\label{ns} && \hspace*{-.5cm} n_{\rm s}=\:
1-6\epsilon_\star\ +\
2\eta_\star\simeq1-{2/ N_\star}=0.964, \>\>\> \\
&& \label{as} \hspace*{-.5cm} \alpha_{\rm s}
=\:{2\over3}\left(4\eta_\star^2-(n_{\rm
s}-1)^2\right)-2\xi_\star\simeq{-2\over N^2_\star}=-6\cdot10^{-4},~~~~~~\\
&& \label{rt} \hspace*{-.5cm} r=16\epsilon_\star\simeq{8/
N_\star}=0.14, \eea\eeqs
where $\xi\simeq m_{\rm P}^4~V'_{\rm I} V'''_{\rm I}/V^2_{\rm I}$
and all the variables with the subscript $\star$ are evaluated at
$\sigma=\sigma_{\star}$. These results are in agreement with the
observational data \cite{wmap, plin,gws} derived in the framework
of the $\Lambda$CDM model.

Since there is no observational hint \cite{plin} for large
non-Gaussianity in the cosmic microwave background, we should make
sure that the masses squared of the scalar excitations in
\Tref{tab}, besides $m^2_\sg$, are greater than the hubble
parameter squared, $\Hhi^2=\Vhi/3\mP^2$, during the last $50-60$
e-foldings of inflation, so that the the observed curvature
perturbation is generated wholly by $\sigma$ as assumed in
\Eref{Prob}. The lowest $m^2_{\chi^\al}$ in \Tref{tab}, by far, is
the one for $\chi^\al=\th$ and its ratio to $\Hhi^2$ is estimated
to be
\beqs\beq {m^2_\th\over\Hhi^2}\lf\sgex\rg\simeq\frac{\kp^2 M^4 \Ns
\lf \Ld^2 (3 - 2\Ns)-12 \ks \rg}{2 \As \Ld^2
\pi^2}+{3\over2\Ns},\label{mH}\eeq
employing \Eref{lan} and under the assumptions made above. Given
that $m^2_\th/\Hhi^2$ increases as $\sg$ drops, we end up with the
following condition:
\beq {m^2_\th/\Hhi^2}\lf\sgex\rg\geq1,\label{m8}\eeq
from which we can derive an upper bound, more restrictive than
that of \Eref{con1}, on $\Ld$
\beq \Ld\lesssim\frac{\sqrt{6|\ks|\Ns}\kp M^2}{\sqrt{\kp^2
M^4\Ns^2+\As \pi^2}}\label{Ld2}\eeq
ranging from $0.74$ to $0.3$ as $\kp$ and $M$ vary from $0.1$ to
$0.01$ -- recall that we use $\ks<0$, as dictated by \Eref{mH}.
The most natural scale close to these $\Ld$ values is the string
scale, i.e., $\Ld=0.1\cdot(5/2.44)\simeq0.2$; we thus confine
ourselves to this choice for $\Ld$ onwards and restrict $\kp$ or
$M$ -- with given $\Ld$. E.g., \Eref{m8} implies:
\beq M \gtrsim\sqrt{\frac{\Ld \pi}{\kp}}\sqrt[4]{\frac{2 \As}{\Ns
\lf \Ld^2 (3 - 2\Ns)-12 \ks \rg}}, \label{kM}\eeq\eeqs
which turns out to be more restrictive than that of \Eref{sgc} if
we make use of \Eref{lan}.

\begin{figure*}[!t]
\centering
\includegraphics[width=60mm,angle=-90]{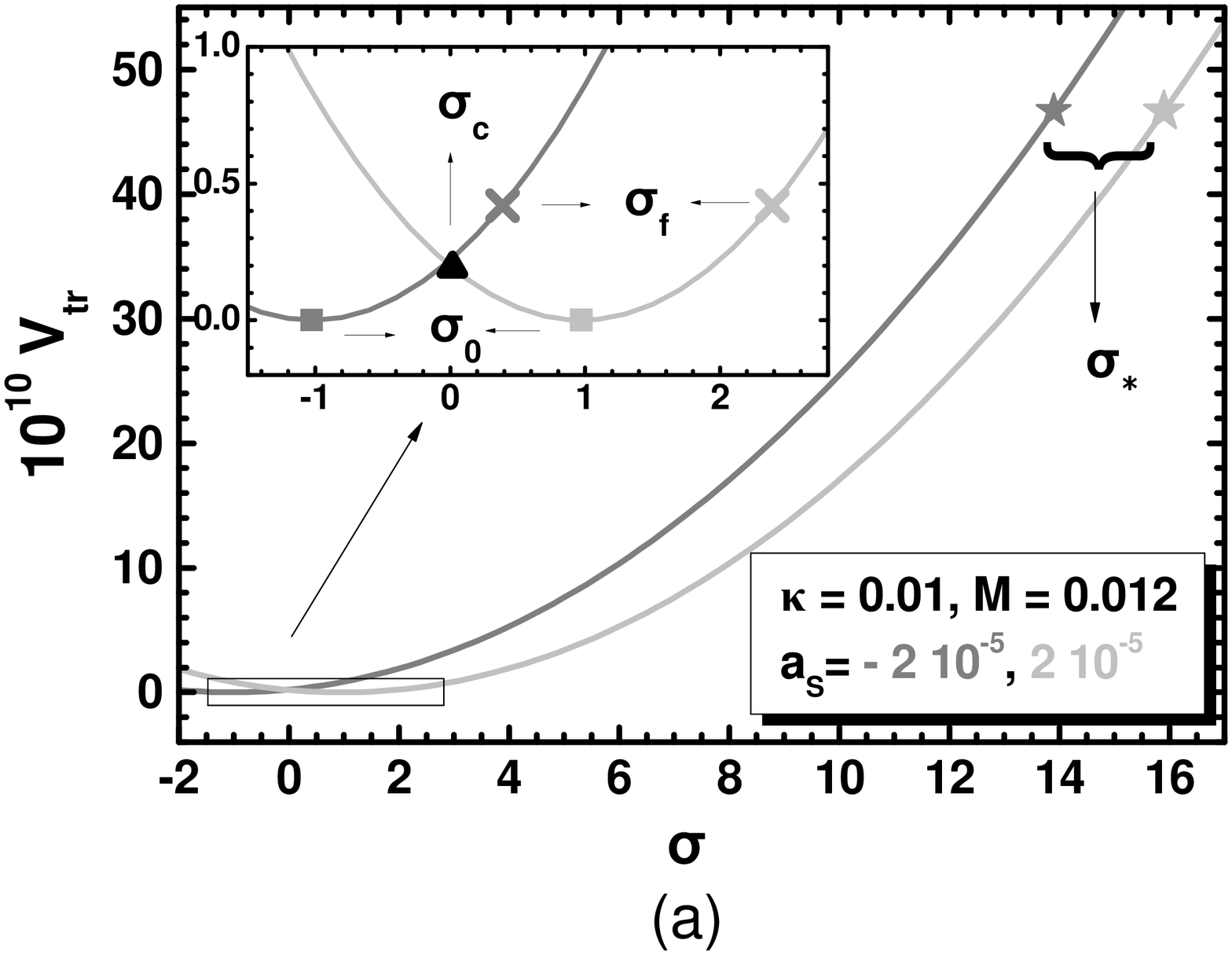}
\includegraphics[width=60mm,angle=-90]{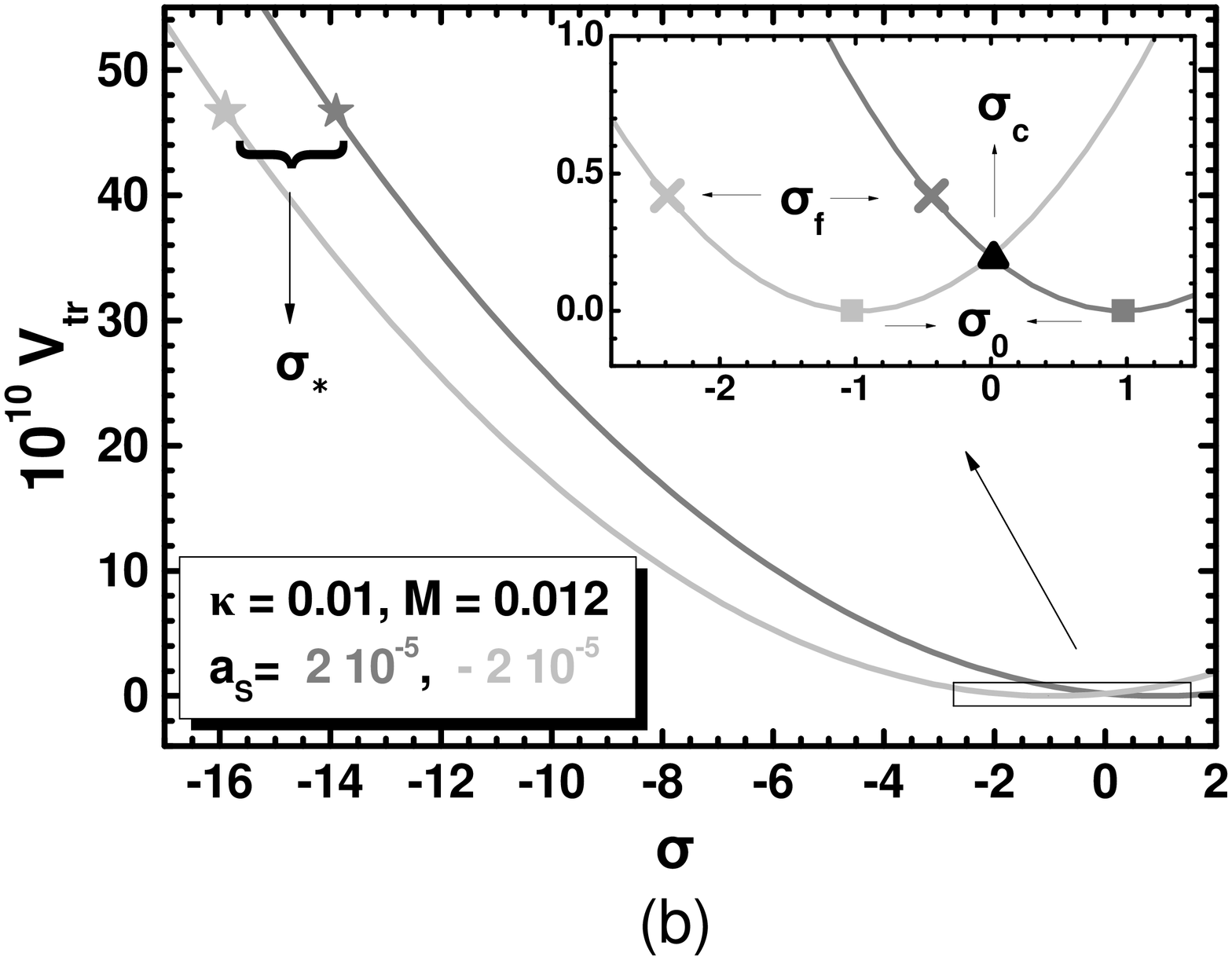}
\caption{\sl Tree level inflationary potential $\Vtr$ as a
function of $\sg$ for $\sgex>0$ and $\aS=-[+]~2\cdot10^{-5}$ (a)
or $\sgex<0$  and $\aS=+[-]~2\cdot10^{-5}$ (b) -- gray [light
gray] line. We set $\kappa=0.01$, $M=0.012,~k_S=-5$ and
$|A_\kp|=10^{-6}$. The values of $\sgex, \sgf, \sg_{0}$ and $\sgc$
are also depicted.}\label{fig1}
\end{figure*}

\subsection{\sf\scshape\small The GUT Phase Transition}

One outstanding feature of our proposal is that the inflationary
scenario is followed by a GUT phase transition, in sharp contrast
to the original quadratic inflation \cite{chaotic}. We should
note, however, that $\Vtr$, \Eref{Vtr}, develops along the track
of \Eref{inftr} an absolute minimum at
\beq \sg_0=\frac{\sqrt{2}\kp\aS M^2}{\ms^2-3\kp^2 M^4},
\label{sg01}\eeq
which has the sign of $\aS$ and a possible complication may be
that $\sg$ gets trapped in this false vacuum and consequently no
GUT phase transition takes place if $\sgc\leq\sg_0$ for $\sgex>0$,
or $\sgc\geq\sg_0$ for $\sgex<0$. Note that the inflationary
observables remain unchanged under the the replacements
\beq \aS\to-\aS~~\mbox{and}~~\sg\to -\sg\,,\label{aSS}\eeq
since $\Vtr$ remains invariant. To assure a timely destabilization
of $\bar \Phi-\Phi$ system -- in the $\phi_{1+}$ or $\phi_{2-}$
direction -- we impose the condition
\beq
\sgc\geq\sg_0~~\mbox{for}~~\sgex>0,~~\mbox{or}~~\sgc\leq\sg_0~~\mbox{for}~~\sgex<0.
\label{sg0}\eeq

The structure of $\Vtr$ for $\sgex >0$ [$\sgex<0$] is visualized
in \sFref{fig1}{a} [\sFref{fig1}{b}], where we present $\Vtr$ --
conveniently normalized such that $\Vtr(\sg_0)=0$ -- as a function
of $\sg$ for the same $\kappa$ and $M$ ($\kappa=0.01$ and
$M=0.012$) and two different $\aS$ values with constant $|\aS|$
taking into account \Eref{Prob}. Namely, in \sFref{fig1}{a}, we
take $\aS=-[+]2\cdot10^{-5}$ -- gray [light gray] line --
corresponding to $\sgex=13.95~[15.9]$ and $\sgf=0.44~[2.4]$. As
anticipated from \eqs{sg0}{sgc}, $\Vtr$ develops minima at the
points $|\sigma_{0}|\simeq0.97$, whereas $\sgc\simeq0.017$ is
constant in all cases since it is independent of $\aS$. We observe
that for $\aS<0$, we obtain $\sigma_{0}<\sgc$ and so the GUT phase
transition can proceed without doubt, whereas for $\aS>0$ we have
$\sigma_{0}>\sgc$, making the destabilization of the $\phi_+$
direction -- see \Tref{tab} -- rather uncertain. In
\sFref{fig1}{b}, we present $\Vtr$ versus $\sg$ changing the signs
of $\aS$ and $\sgex$ according \Eref{aSS}, i.e., we set
$\aS=+[-]2\cdot10^{-5}$ with $\sgex=-13.95~[-15.9]$ -- gray [light
gray] line. We remark that the case with $\aS<0$ remains
problematic since $\sg$ meets first $\sg_0=-0.97<\sgc=0.017$ and
its trapping in the minimum is possible, whereas the case with
$\aS>0$ is free from such a problem, since
$\sg_0=0.97>\sgc=0.017$. Given this situation we henceforth
concentrate on the case with $\sgex>0$. The results for the case
with $\sgex<0$ are obtained by flipping the sign of $\aS$ as
suggested by the symmetry of $\Vtr$, \Eref{aSS}.

\subsection{\sf\scshape\small Compatibility With the Formation of
Cosmic Strings}

If $\Ggut=G_{B-L}$, $B-L$ cosmic strings are produced during the
GUT phase transition, at the end of inflation. The tension
$\mu_{\rm cs}$ of these defects has to respect the bound
\cite{jp,mfhi, plcs}:
\begin{equation} \label{mucs} \mu_{\rm cs}=
{9.6\pi
M^2\over\ln(2/\beta)}\leq8\cdot10^{-6}~~\Rightarrow~~M\leq0.001\lf{\ln(2/\beta)\over1.2\pi}\rg^{1/2},\end{equation}
where $\beta=\kappa^2/8g^2\leq10^{-2}$ with $g\simeq0.7$ being the
gauge coupling constant close to $\Mgut$. From \Eref{mucs}, for
$\kp=0.1,0.01$ and $0.001$, we obtain $10^{3}M\leq1.33,1.7$ and
$2$, whereas \Eref{kM} entails $10^{3}M\geq2.7,8.7$ and $27$
respectively. As a consequence, our scheme is not compatible with
the choice $\Ggut=G_{B-L}$. This negative result can be, most
probably, avoided if we invoke the superpotential employed in
shifted \cite{shifted} or smooth \cite{smooth} FHI. In that cases,
$\Phi$ and $\bar\Phi$ are confined to some non-vanishing value
during inflation; thus, the $B-L$ strings can be easily inflated
away.

\section{Results}\label{res}

Following our previous discussion we henceforth concentrate our
analysis on $\Ggut=\Glr$ or $\Gfl$. For both selected $G$'s, $M$
can be related to the GUT scale since the non-singlet under $\Gsm$
gauge bosons acquire mass equal to $gM$ at the SUSY vacuum,
\Eref{vevs} -- see \cref{rlarge}. However, in high-scale SUSY
\cite{strumia, ibanez} the GUT scale is model dependent and so any
$M$ value between $0.001$ and $0.1$ is, in principle, acceptable.
For reference we mention that the conventional SUSY GUT scale
corresponds to the choice $gM=(2/2.44)\cdot10^{-2}$, i.e.,
$M\simeq0.012$. Recall finally that we set $\ks=-\ksp=-5$
$|A_\kp|=10^{-6}$ and $\Ld=0.2$ throughout.

In our numerical calculations, we use the complete formulae for
$\Vhi$, $\Ns$, $\As$ and the slow-roll parameters -- see
\eqss{Vhic}{Nhi}{slow} -- and not the approximate relations listed
in \Sref{fhi4} for the sake of presentation. As regards $Q$ in
\Eref{Vrc}, we determine it by requiring \cite{Qenq} $\Delta
V(\sgex)=0$. Note that $Q$ is not well-defined if we impose the
alternative condition \cite{Qenq} $\Delta V(\sgf)=0$ since
$m_{\phi+}$ instantaneously vanishes when $\sgf=\sgc$. To reduce
the possible \cite{circ, Qenq} dependence of our results on the
choice of $Q$, we confine ourselves to values of $\kp, M$ and
$\aS$ which do not enhance $\Delta V$. As a consequence, our
findings are highly independent of the specific choice of $G$. For
definiteness we mention that we take $G=\Gfl$.

\begin{figure*}[!t]
\centering
\includegraphics[width=60mm,angle=-90]{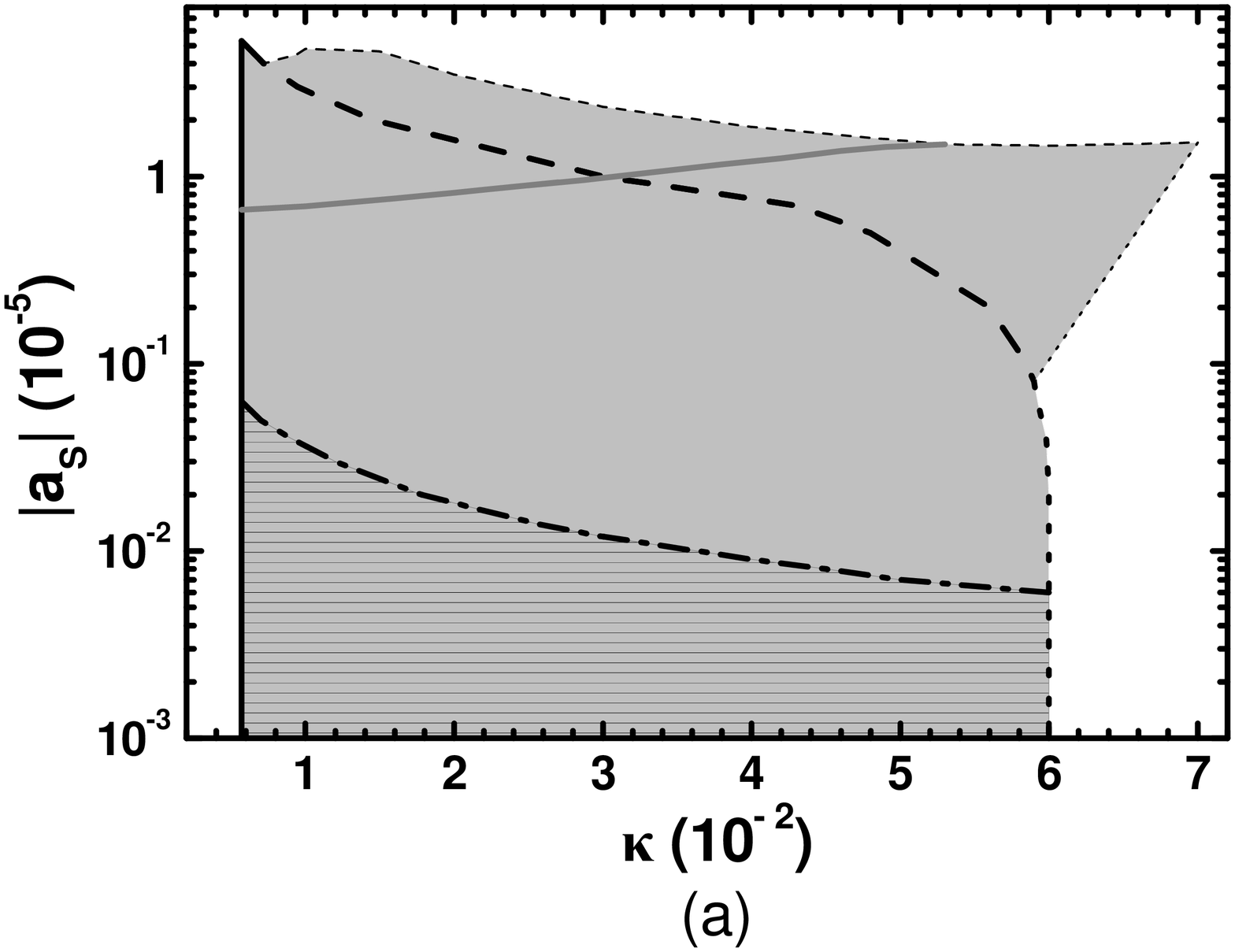}
\includegraphics[width=60mm,angle=-90]{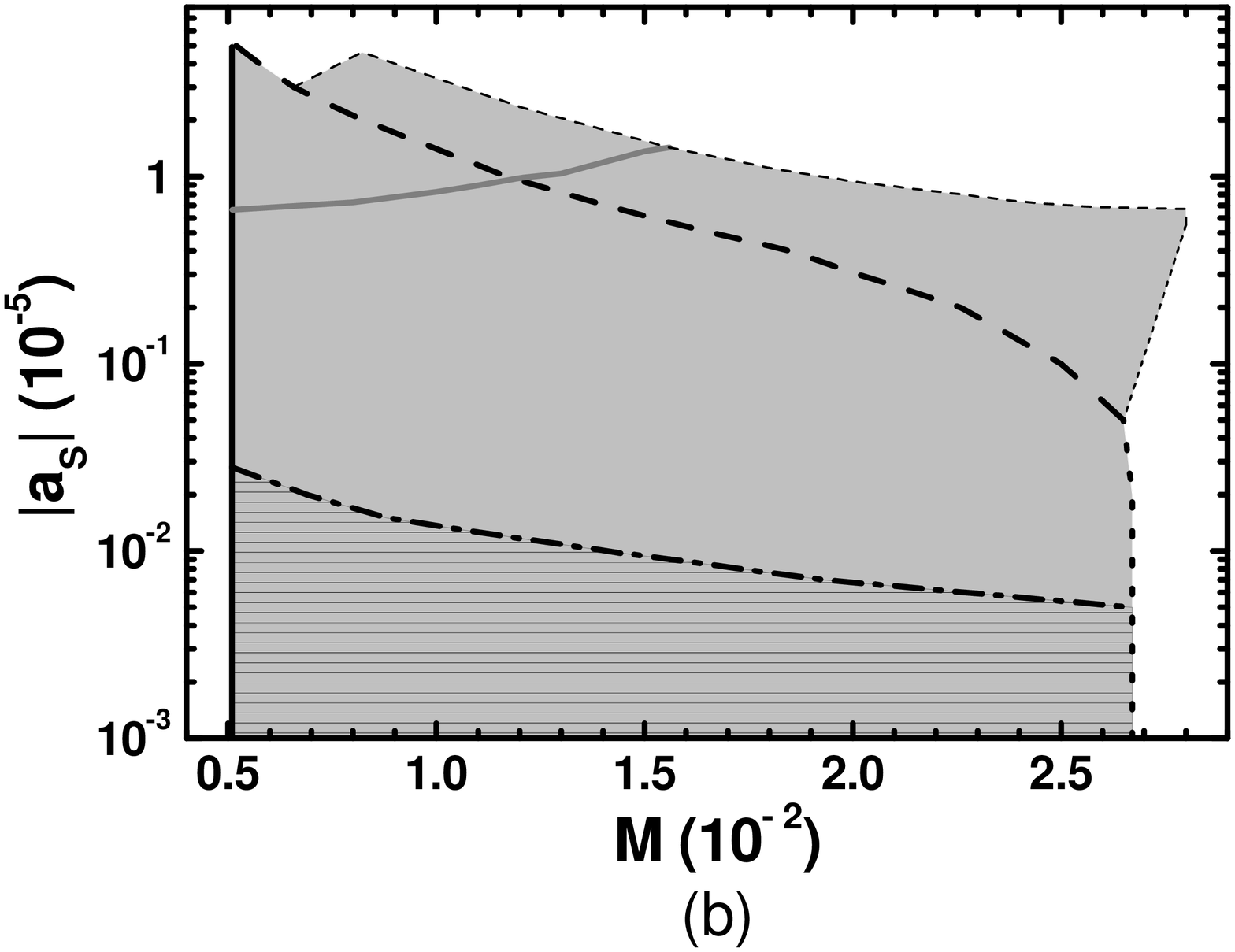}
\caption{\label{fig2}\sl Allowed (shaded [lined]) regions for
$\sgex>0$ and $\aS<0$ [$\aS>0$] in the $\kappa-|\aS|$ plane with
$M=0.012$ (a) and in the $M-|\aS|$ plane with $\kp=0.03$ (b).
Along the gray line we set $\aS=-\ms$. We take $k_S=-\ksp=-5$,
$|A_\kp|=10^{-6}$ and $\Ld=0.2$. }
\end{figure*}

Confronting our model with the imposed constraints, we depict the
allowed (lightly gray shaded [lined]) regions for $\aS<0$
[$\aS>0$] in the $\kappa-|\aS|$ plane with $M=0.012$ and in the
$M-|\aS|$ plane with $\kp=0.03$ -- see \sFref{fig2}{a} and
{\sf\ftn (b)} respectively. The left bounds in both plots come
from the saturation of \Eref{m8}. It is straightforward to show
that the (simplified) analytical expression in \Eref{kM} is in
accordance with the bound, $0.0057$ [$0.0051$] depicted in
\sFref{fig2}{a} [\sFref{fig2}{b}]. Had we used $\ks=-1$, this
bound in \sFref{fig2}{a} [\sFref{fig2}{b}] would have been moved
to $0.014$~[$0.008$] cutting a minor slice of the allowed region.
It is clear from \Eref{sg0} that the allowed region for $\aS>0$ is
considerably shrunk compared to that for $\aS<0$, since $\aS<0$
implies $\sg_0<0$, and so \Eref{sg0} is automatically fulfilled
thanks to the positivity of $\sgc$ -- see \Eref{sgc}. Indeed, the
saturation of \Eref{sg0} gives the upper bound of the allowed
(lined) regions for $\aS<0$. On the other hand, for $\aS<0$ no
solution to \Eref{Prob} exists beyond the thin dashed line. In the
shaded region between the thick and thin dashed lines the end of
inflation is found by the condition $\sgf=\sgc$ and not the one in
\Eref{slc} which exclusively gives $\sgf$ for $\aS>0$, and in the
regions below the thick dashed lines for $\aS<0$. Note that for
$\aS<0$ we have allowed parameters even for $|\aS|=\ms$ which are
depicted by the gray lines. Finally, beyond the (thin and thick)
dotted lines, our results become unstable with respect to the
variations of $Q$; the model predictions are, thus, less trustable
and we do not pursue it any further.

Summarizing our findings from \Fref{fig2} the parameters of $W$ in
\Eref{Whi} are bounded as follows:
\beq0.57\lesssim\kp/10^{-2} \lesssim7~~\mbox{and}~~0.51\lesssim
M/10^{-2} \lesssim2.9\,.\label{res1}\eeq
Moreover, the SSB mass parameters in \Eref{Vis0} are confined in
the following ranges:
\beq0.66\lesssim\ms/10^{-5} \lesssim4.4~~\mbox{and}~~
|\aS|/10^{-5} \lesssim 5.3 ~~[0.063]\label{res2}\eeq
for $\aS<0$ [$\aS>0$]. The most natural framework of SSB in which
our model can be embedded is that of high-scale SUSY since the
$\ms$ values encountered here are roughly consistent with
$m_h\simeq126~\GeV$ \cite{ibanez}. On the other hand, split SUSY
cannot be directly combined with our proposal since requiring
$m_h\simeq126~\GeV$ implies \cite{strumia} $\ms\leq10^8~\GeV$,
which is rather low to drive inflation. However, a possible
coupling of $S$ with the electroweak higgses of the minimal SUSY
SM can modify this conclusion as outlined in \cref{strumia}.

It is worth noticing that, contrary to \cref{rlarge1}, $\kp$ and
$M$ are constrained so that the contribution to $\Vhi$ from
\Eref{Vis0} exceeds that from \Eref{Vhi0}. As a consequence, our
model here shares identical predictions with the original
quadratic inflationary model as regards $\ns,\as$ and $r$, and so
it is consistent with \bicep\ findings \cite{gws}. Indeed, for
$\Ns=55$ we find $0.12\lesssim r\lesssim0.14$ and
\beq0.963\lesssim \ns\lesssim0.969,~~~~4.7\lesssim -\as/10^{-4}
\lesssim6.8\label{res3}\eeq
which are consistent with WMAP \cite{wmap} and Planck \cite{plin}
results within the $\Ld$CDM model. Contrary to quadratic model,
however, our model implies a built-in mechanism for spontaneous
breaking of $G$ at the scale $M$, \Eref{res1}, compatible with the
SUSY GUT scale, $M\simeq0.012$. The resulting mass of the inflaton
at the SUSY vacuum takes values
\beq 6.5\lesssim m_\sg/10^{-6}\lesssim8.7, \label{res4}\eeq
which allow for the decay of the inflaton to right-handed
neutrinos, if the relevant couplings exist. Thus, a successful
scenario of non-thermal leptogenesis, along the lines of
\cref{lept,mfhi}, can be easily constructed. 

\section{Conclusions}\label{con}

We have presented a framework for implementing quadratic (chaotic)
inflation in realistic SUSY models which have previously been used
for FHI. Namely, we have retained a $U(1)$ R-symmetry from earlier
FHI which yields a unique superpotential, $W$, at renormalizable
level, linear with respect the inflaton field. On the other hand,
the \Kap, $K$, is judiciously chosen so that no extensive SUGRA
corrections arise. Our model is thus protected against
contributions from higher order terms in both $K$ and $W$. We
showed that the model displays a wide and natural range of the
parameters $\kp, M$ and $\aS$ which allows quadratic inflation to
be successfully implemented, provided that the SSB mass parameter
$\ms$ lies at the intermediate energy scale motivated by
high-scale (or, under some special circumstances, split) SUSY
breaking. As a consequence the inflationary observables are in
excellent agreement with the combined analysis of the Planck, WMAP
and \bicep\ measurements.


\acknowledgments Q.S. thanks Gia Dvali, Ilia Gogoladze, Matt
Civiletti, and Tianjun Li for helpful discussions and acknowledges
support by the DOE grant No. DE-FG02-12ER41808. C.P. acknowledges
support from the Generalitat Valenciana under grant
PROMETEOII/2013/017.


\def\ijmp#1#2#3{{\sl Int. Jour. Mod. Phys.}
{\bf #1},~#3~(#2)}
\def\plb#1#2#3{{\sl Phys. Lett. B }{\bf #1}, #3 (#2)}
\def\prl#1#2#3{{\sl Phys. Rev. Lett.}
{\bf #1},~#3~(#2)}
\def\rmp#1#2#3{{Rev. Mod. Phys.}
{\bf #1},~#3~(#2)}
\def\prep#1#2#3{{\sl Phys. Rep. }{\bf #1}, #3 (#2)}
\def\prd#1#2#3{{\sl Phys. Rev. D }{\bf #1}, #3 (#2)}
\def\npb#1#2#3{{\sl Nucl. Phys. }{\bf B#1}, #3 (#2)}
\def\npps#1#2#3{{Nucl. Phys. B (Proc. Sup.)}
{\bf #1},~#3~(#2)}
\def\mpl#1#2#3{{Mod. Phys. Lett.}
{\bf #1},~#3~(#2)}
\def\jetp#1#2#3{{JETP Lett. }{\bf #1}, #3 (#2)}
\def\app#1#2#3{{Acta Phys. Polon.}
{\bf #1},~#3~(#2)}
\def\ptp#1#2#3{{Prog. Theor. Phys.}
{\bf #1},~#3~(#2)}
\def\n#1#2#3{{Nature }{\bf #1},~#3~(#2)}
\def\apj#1#2#3{{Astrophys. J.}
{\bf #1},~#3~(#2)}
\def\mnras#1#2#3{{MNRAS }{\bf #1},~#3~(#2)}
\def\grg#1#2#3{{Gen. Rel. Grav.}
{\bf #1},~#3~(#2)}
\def\s#1#2#3{{Science }{\bf #1},~#3~(#2)}
\def\ibid#1#2#3{{\it ibid. }{\bf #1},~#3~(#2)}
\def\cpc#1#2#3{{Comput. Phys. Commun.}
{\bf #1},~#3~(#2)}
\def\astp#1#2#3{{Astropart. Phys.}
{\bf #1},~#3~(#2)}
\def\epjc#1#2#3{{Eur. Phys. J. C}
{\bf #1},~#3~(#2)}
\def\jhep#1#2#3{{\sl J. High Energy Phys.}
{\bf #1}, #3 (#2)}
\newcommand\jcap[3]{{\sl J.\ Cosmol.\ Astropart.\ Phys.\ }{\bf #1}, #3 (#2)}
\newcommand\njp[3]{{\sl New.\ J.\ Phys.\ }{\bf #1}, #3 (#2)}

\end{document}